\documentclass{article}
\usepackage{spconf,amsmath,graphicx}
\usepackage{enumitem}
\usepackage{siunitx}
\usepackage{booktabs}
\usepackage{tablefootnote}
\usepackage[hang,flushmargin]{footmisc}

\title{Wav2CLIP: Learning Robust Audio Representations From CLIP}
\name{\begin{tabular}{c}Ho-Hsiang Wu$^{1, \star}$, Prem Seetharaman$^{2}$, Kundan Kumar$^{2}$, Juan Pablo Bello$^{1}$\end{tabular}
\thanks{$^{\star}$ Work done during an internship at Descript, Inc.}
}
\address{$^{1}$ Music and Audio Research Laboratory, New York University, USA \\
         $^{2}$ Descript, Inc.}

\begin{document}
\ninept
\maketitle
\begin{abstract}

We propose Wav2CLIP, a robust audio representation learning method by distilling from Contrastive Language-Image Pre-training (CLIP). We systematically evaluate Wav2CLIP on a variety of audio tasks including classification, retrieval, and generation, and show that Wav2CLIP can outperform several publicly available pre-trained audio representation algorithms. Wav2CLIP projects audio into a shared embedding space with images and text, which enables multimodal applications such as zero-shot classification, and cross-modal retrieval. Furthermore, Wav2CLIP needs just $\sim$10\% of the data to achieve competitive performance on downstream tasks compared with fully supervised models, and is more efficient to pre-train than competing methods as it does not require learning a visual model in concert with an auditory model. Finally, we demonstrate image generation from Wav2CLIP as qualitative assessment of the shared embedding space. Our code and model weights are open sourced and made available for further applications.

\end{abstract}
%
\section{Introduction}
\label{sec:intro}


Self-supervised learning is a highly active and fruitful area of research, yielding state-of-the-art (SOTA) results in speech recognition \cite{oord2018representation,schneider2019wav2vec,ravanelli2020multi}, music classification \cite{spijkervet2021contrastive,wu2021multi}, computer vision \cite{chen2020simple,kolesnikov2020big}, and natural language processing \cite{devlin2018bert,brown2020language}. Through self-supervised learning, one can learn
robust representations by exploiting the intrinsic structure of data, from a much larger amount of data than possible through supervised learning, where human annotation of large amounts of data is very expensive. Once models are pre-trained, they can be transferred via fine-tuning to solve tasks with less data. 

\begin{figure}[t]
\centering
\includegraphics[width=\linewidth]{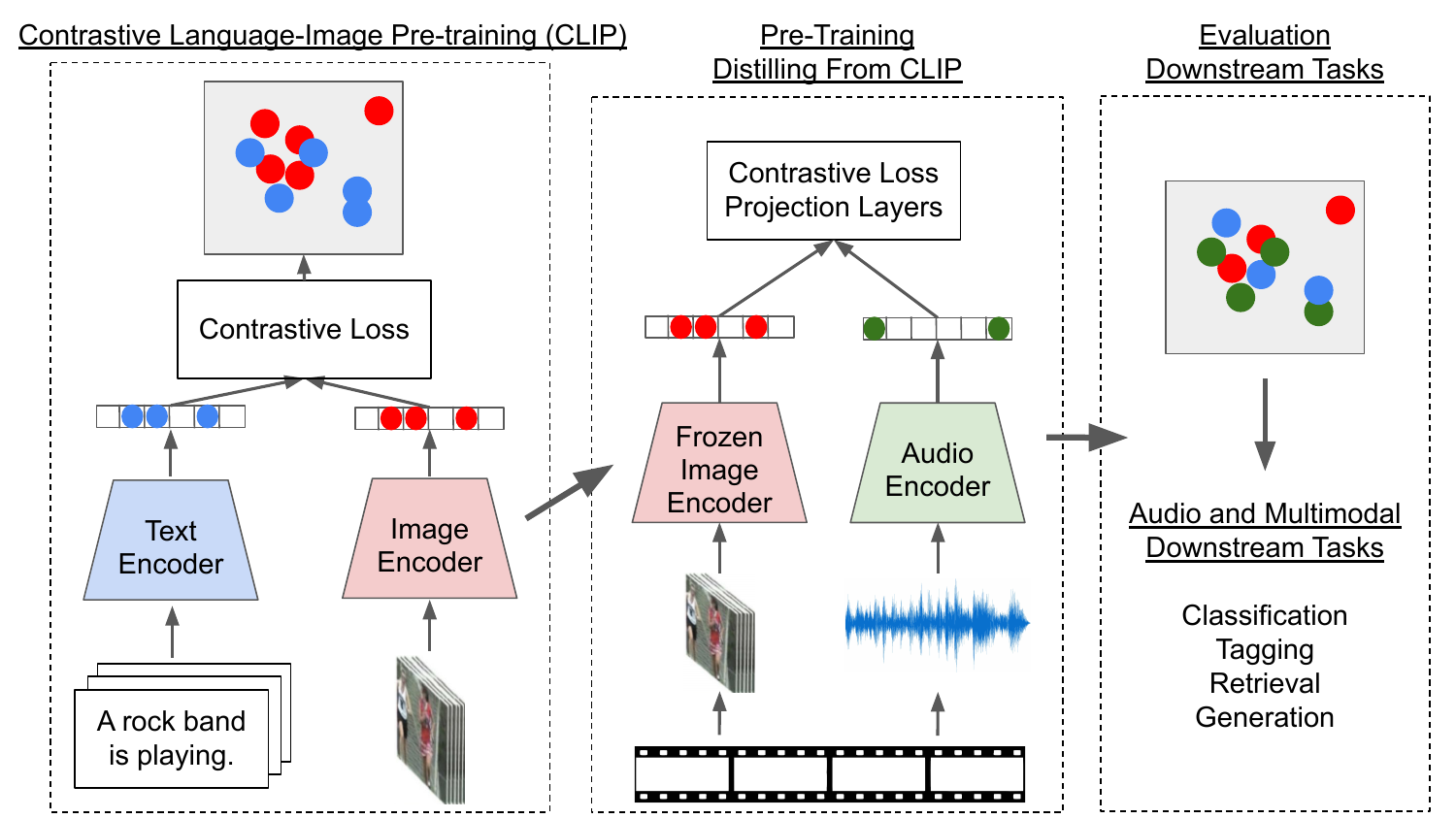}
\caption{Contrastive Language-Image Pre-training (CLIP), and our two-stage approaches including pre-training and evaluation.} 
\label{fig:method}
\vspace{-1.0em}
\end{figure}

A highly successful form of self-supervised learning is via exploiting \textit{multimodal correspondence}. For example, in audio, predicting correspondence between audio and visual streams inherent in large collections of video (e.g. YouTube) results in robust audio models that can be applied to downstream audio tasks \cite{aytar2016soundnet,arandjelovic2017look,alwassel2019self,cramer2019look,tzinis2020into}. Text and image correspondence has also been leveraged for self-supervised learning \cite{chen2020uniter}, including a recently proposed powerful approach: \textbf{Contrastive Language–Image Pre-training (CLIP)} \cite{radford2021learning}. CLIP exhibits SOTA results in zero-shot image classification, image retrieval via text, and even guiding generative models. It does so by leveraging a large number of image/text pairs that are scraped from the internet, and training a model via a contrastive loss which predicts if a text and an image pair either corresponds with one another or does not as in Figure \ref{fig:method}. Through this simple training objective, CLIP learns highly generalizable representations in both domains, and enables tasks like image classification, image captioning, cross-modal image retrieval with text prompts, and more.

In this work, we propose Wav2CLIP, an audio-visual correspondence model trained by distilling from the CLIP model. Wav2CLIP works by freezing the CLIP vision model, running it on the visual streams from videos, and training a new model to predict the same CLIP embeddings from the audio stream alone. Our method is simple to train, extensible, and outputs embeddings that are aligned with images. Unlike previous audio-visual correspondence models \cite{cramer2019look}, it is not necessary to learn a visual model jointly with the auditory model, making the Wav2CLIP training recipe very lightweight. Further, Wav2CLIP embeddings are derived from CLIP, which means they are aligned with text. This makes zero-shot audio classification possible, as well as audio captioning, and cross-modal text/audio-to-audio retrieval, which has applications in various domains (e.g. searching for the sound of glass breaking in surveillance videos, or finding the perfect Foley sound effect when editing sound for a movie). Finally, Wav2CLIP is easy to scale to large video datasets.

We systematically evaluate Wav2CLIP representations across a wide variety of audio tasks, including classification and retrieval, and compare with other audio representation learning approaches, as well as SOTA results from each task. We show how to apply Wav2CLIP to solve several multimodal and zero-shot tasks. Finally, we use Wav2CLIP to guide a generative model - VQGAN-CLIP\footnote{https://github.com/nerdyrodent/VQGAN-CLIP} - directly from audio, validating that the embedding space is meaningful. We also note that there is concurrent work - AudioCLIP \cite{guzhov2021audioclip}. In contrast to their work, we do not learn the visual encoder but rather distill CLIP into an audio model, resulting in one joint embedding space for three different modalities. For reproducibility and potential applications, we open source code and model weights\footnote{https://github.com/descriptinc/lyrebird-wav2clip}.

\section{Method}
\label{sec:method}

Contrastive Language–Image Pre-training (CLIP) \cite{radford2021learning} is trained with massive image and text pairs from the internet with noise-contrastive estimation \cite{gutmann2010noise}, where image and text pairs from the same sample is used as positive examples, and all the others in the same batch are treated as negatives. Both CLIP image and text representations are projected to a shared embedding space as shown in Figure \ref{fig:method} on the left, with the key idea that natural language can be leveraged as a flexible prediction space to help generalize and transfer knowledge, and the continuous conceptual space of images is mapped to the discrete symbolic space of text. As a result, CLIP has shown great success applied to tasks such as zero-shot classification and cross-modal retrieval, and has also been extended to cross-modal generation \cite{ramesh2021zero}.




We take a two-stage approach as shown in Figure \ref{fig:method}. First, our pretext task is to pre-train audio encoders by distilling CLIP\footnote{https://github.com/openai/CLIP} image embeddings through videos. We follow the original CLIP paper by using a contrastive loss \cite{oord2018representation}, and experiment with adding multi-layer perceptron (MLP) as projection layers, similar to \cite{chen2020simple,grill2020bootstrap,recasens2021broaden}. Heuristically, we found that loss function with cross projection defined as $CX Loss = L(f(Image), Audio) + L(Image, g(Audio))$ ($f, g$: projection functions and $L$: contrastive loss), was helpful for stabilizing the distillation procedure. We hypothesize it is because CX Loss provides additional trainable MLP layers for more flexibility. It also enforces multimodal consistency, in that the learned audio embedding can be used to recover the CLIP image embedding through a projection layer. Throughout distillation, the original CLIP model weights are kept frozen. After the audio encoder is pre-trained, it is used as a frozen feature extractor, and additional layers can be trained to solve downstream tasks.

\section{Experimental Design}
\label{sec:experimental_design}




\subsection{Pre-training: Distilling From CLIP}
\label{sssec:pretraining}


Throughout all experiments, we use ResNet-18 architecture as audio encoder, as implemented in the VGGSound \cite{chen2020vggsound} baseline model\footnote{https://github.com/hche11/VGGSound}. For the audio encoder, we first transform the raw audio waveform (1D) into spectrogram (2D) as input to the ResNet, and take average pooling to output one embedding for the entire audio sequence with 512 dimensions. For the tasks which require frame-level embeddings, such as audio retrieval and audio captioning, we first take segments of input audio with 1s non-overlapping window in order to output frame-level embeddings along time axis. This frame-level adaptation only occurs at inference time. We pre-trained with CX Loss on VGGSound, an audio-visual YouTube video dataset containing $\sim$200k 10-second clips (16kHz sample rate) labeled with 309 multi-classes. Note that we did not use the annotated labels during pre-training. We randomly sampled 5-second video from training split, extract CLIP image embeddings for each frame (30 fps, resulting in 30$\times$5$=$150 frames) and use mean pooling to get clip-level embeddings. We use Adam optimizer with learning rate 0.001, with standard learning rate reduce on plateau, and early stopping.



\subsection{Evaluation: Downstream Tasks}
\label{sssec:downstream}

After pre-training, we freeze the learned weights and use the audio encoder as a feature extractor for all downstream tasks. We also train supervised learning classifiers from scratch for each downstream classification dataset as reference. We select diverse set of data ranging from various number of clips, number of categories, and perform diverse tasks including classification, retrieval, and generation. For evaluation, we use relevant metrics detailed in Table \ref{tab:audio_tasks} for each task.

\subsubsection{Dataset and Tasks}

We select various size of publicly available audio classification datasets, which are commonly used for evaluation \cite{cramer2019look}, as well as several audio tasks/datasets from Detection and Classification of Acoustic Scenes and Events (DCASE)\footnote{http://dcase.community} challenges, as shown in Table \ref{tab:audio_tasks} including classification, retrieval, and audio captioning. ESC-50 \cite{piczak2015dataset} is a relatively simple dataset with only 2k samples, UrbanSound8K \cite{Salamon:UrbanSound:ACMMM:14} is a larger environmental dataset with 10 classes. FSD50K \cite{fonseca2020fsd50k} is 5x larger than UrbanSound8K, and is a more challenging dataset with multi-labels. VGGSound \cite{chen2020vggsound} is a very large audio-visual dataset with the most and diverse general sound classes among all. TAU \cite{9415085} is another audio-visual dataset focused on background acoustic scene, compared to the different type of foreground sound events from VGGSound. DESED is re-purposed as audio retrieval (AR) task, as it provides segment-level annotations which can be used for audio segment retrieval within a clip. Finally, Clotho \cite{Drossos_2020_icassp} is a unique dataset for audio captioning (AC) task.

\begin{table}[htb]
\centering
\begin{tabular}{@{}ccl@{}r@{}c@{}}\toprule
Dataset & Task & \# Clip (Split) & \# Class & Metric \\
\midrule
ESC-50 \cite{piczak2015dataset} & MC/ZS & 2k (5 folds) & 50 & ACC \\
UrbanSound8K \cite{Salamon:UrbanSound:ACMMM:14} & MC/ZS & 8k (10 folds) & 10 & ACC \\
FSD50K \cite{fonseca2020fsd50k} & ML/ZS & 50k & 200 & mAP \\
VGGSound \cite{chen2020vggsound} & MC/ZS & 185k & 309 & mAP \\
TAU Audio \cite{9415085} & MC/ZS & 190k & 10 & ACC \\
\midrule
DESED \cite{Turpault2019_DCASE} & AR & 2.5k (valid) & 10 & F1 \\
VGGSound \cite{chen2020vggsound} & CMR & 15k (test) & 309 & MRR \\
\midrule
Clotho \cite{Drossos_2020_icassp} & AC & 5k (evaluation) & & COCO\tablefootnote{\label{coco}https://github.com/tylin/coco-caption} \\
\bottomrule
\end{tabular}
\caption{Downstream tasks/datasets, including 1. classification: multi-class (MC), multi-label (ML), zero-shot (ZS), 2. retrieval: audio (AR) and cross-modal retrieval (CMR), and 3. audio captioning (AC) task, with various \# of clips, \# of classes, and common metrics.}
\label{tab:audio_tasks}
\vspace{-1.0em}
\end{table}

\begin{table*}[htb]
\centering
\begin{tabular}{@{}c@{\hskip 0.1in}c@{\hskip 0.1in}c@{\hskip 0.1in}c@{\hskip 0.1in}c@{\hskip 0.1in}c@{\hskip 0.1in}c@{\hskip 0.1in}c@{\hskip 0.1in}c@{}}\toprule
& \multicolumn{5}{c}{Classification} & \multicolumn{3}{c}{Retrieval} \\
\cmidrule(lr){2-6} \cmidrule(lr){7-9}
Model & ESC-50 & UrbanSound8K & FSD50K & VGGSound & TAU Audio Only & DESED (AR) & \multicolumn{2}{c}{VGGSound (CMR)} \\
\cmidrule(lr){8-9}
& ACC & ACC & mAP & mAP & ACC & F1 & A$\rightarrow$I (MRR) & I$\rightarrow$A (MRR) \\
\midrule
Supervise & 0.5300 & 0.6289 & 0.3212 & 0.4512 & 0.3827 \\
OpenL3 & 0.723 & 0.7681 & 0.4053 & 0.3405 & \textbf{0.651} \cite{9415085} & 0.1176 & 0.0171 & 0.0160 \\
YamNet & 0.8505 & 0.7832 &  \textbf{0.5039} & 0.4541 & 0.5667 & 0.3420 \\
Wav2CLIP & \textbf{0.8595} & \textbf{0.8101} & 0.4308 & \textbf{0.4663} &  0.6302 & 0.3955 & 0.0566 & 0.0678 \\
\midrule
SOTA & 0.959 \cite{guzhov2021audioclip} & 0.8949 \cite{guzhov2021audioclip} &  0.5671 \cite{gong2021psla} & 0.544 \cite{Kazakos2021SlowFastAuditory} & 0.687 \cite{Fedorishin2021} \\
\midrule
Wav2CLIP (ZS) & 0.414 & 0.4044 & 0.0302 & 0.1002 & 0.1726 \\
\bottomrule
\end{tabular}
\caption{Results of downstream classification and retrieval tasks, compared with supervised training of each task from scratch, other audio representation models - OpenL3, YamNet - and current state-of-the-art (SOTA). ZS indicates using Wav2CLIP as a zero-shot model.}
\label{tab:all_tasks}
\vspace{-1.0em}
\end{table*}

For multi-class (MC) classification tasks, we train a simple 2-layer MLP classifier with corresponding number of classes as output. For multi-label (ML) classification, the softmax layer is replaced with BCEWithLogitsLoss optimized for multi-label predictions. If the folds are not provided, we run 3 trials for statistical significance. We report standard metrics that are also used in other benchmarks.

For audio retrieval (AR) on DESED, we first extract frame-level (w/ 1s non-overlapping window) embeddings from the validation split. Then all labeled segments are used as query to retrieve all other segments with the same labels as predictions. The validation set is randomly split into 5 folds, and we use 1 hold-out cross validation to pick decision threshold (in order to determine search hit from cosine distances), with 0.1s time\_resolution on segmented-based metrics. We report F1 metric using sed\_eval\footnote{https://github.com/TUT-ARG/sed\_eval}.

We also explore the capability for Wav2CLIP to solve multimodal tasks, and demonstrate how to zero-shot transfer to other modalities. Wav2CLIP audio embeddings are projected to the shared embedding space as CLIP, therefore, we get image and text modality for free. For cross-modal retrieval (CMR) and zero-shot classification (ZS), we first extract CLIP image and Wav2CLIP audio embeddings from VGGSound test split, CLIP text and Wav2CLIP audio embeddings from labels and audio of classification dataset, we compute cosine distance between these image/text and audio embeddings, in order to retrieve relevant clips for cross-modal retrieval, and assign labels to the closet text for classification. We report mean reciprocal rank (MRR) on CMR where clips with the same labels are treated as relevant. For audio captioning (AC), we freeze our audio encoder as feature extractor and only train a 1-layer transformer decoder to predict text sequences. We follow the DCASE challenge setup\footnote{\label{clotho}http://dcase.community/challenge2020/task-automatic-audio-captioning} and report standard COCO caption evaluation\footnotemark[6] metrics. 




\subsection{Baseline Comparisons}
First, we train from scratch (randomly initialize encoder weights) directly on each downstream dataset as Supervise baseline. Next, we compare Wav2CLIP with other publicly available audio representation algorithms, OpenL3 \cite{cramer2019look}, and YamNet\footnote{https://github.com/tensorflow/models/tree/master/research/audioset/yamnet}, as these two algorithms are pre-trained with completely different pretext tasks. OpenL3 is trained with multimodal self-supervision on AudioSet \cite{45857}. YamNet is also trained on AudioSet, but via supervision from annotated labels with only audio. Both OpenL3 and YamNet are commonly referred to as strong baselines for performance comparisons on various audio tasks, such as audio classification and retrieval. We extract OpenL3 (512 dim), YamNet (1024 dim), and Wav2CLIP (512 dim) features, applying same training recipes to all downstream classification and retrieval tasks. YamNet and OpenL3 both output frame-level embeddings, we take mean pooling across time for clip-level embeddings. We also include SOTA results from the literature to the best of our knowledge, for topline results specifically trained for each task/dataset as upper bound compared to our approach.

To understand better the amount of annotated data required for each pre-training model, we also take VGGSound, the largest annotated dataset in our experiments, and evaluate on classifiers trained with 1, 2, 5, 10, 20, 50, and 100\% of training split.

\section{Results and Discussions}
\label{sec:results}



\subsection{Classification, Retrieval, and Audio Captioning Tasks}


From Table \ref{tab:all_tasks}, we see that Supervise baseline does not perform well across all classification tasks, except for VGGSound, where massive annotated data is available. This indicates that, pre-training in general helps, especially for those tasks with limited labeled data. Next, we compare Wav2CLIP with other publicly available audio representations, Wav2CLIP performs slightly better or similarly to YamNet and OpenL3 in almost all tasks, except for FSD50K, where there is still a gap. Our hypothesis is that YamNet is pre-trained supervisedly with AudioSet, which is also a multi-label dataset with the same label taxonomy as superset of FSD50K, therefore the pretext task matches very closely to the downstream task. We also observe similar behavior on TAU, an audio-visual dataset, both OpenL3 and Wav2CLIP, pre-trained with audio-visual self-supervision perform better compared to YamNet, only trained with audio modality.

When compared with SOTA, there is still room for improvement, since for most of the SOTA models the encoders are trained or fine-tuned on each task specifically, while Wav2CLIP is used as frozen feature extractor and only simple MLP classifiers are trained to output corresponding number of classes, therefore, only one version of audio encoder is required for all tasks. On the last row of Table \ref{tab:all_tasks}, we see all zero-shot (ZS) classification results, which come for free without any extra training. It does not perform as good as MLP classifiers, but it is not random, especially for ESC-50 and UrbanSound8K. We think that the more classes, the more challenging for zero-shot as it is more difficult to maintain clear boundary between each class label in the existing embedding space without further fine-tuning, this is especially worse for FSD50K, as multi-label brings more complexity.


\begin{table}[htb]
\centering
\begin{tabular}{@{}cc@{\hskip 0.075in}c@{\hskip 0.075in}c@{\hskip 0.075in}c@{\hskip 0.075in}c@{\hskip 0.075in}c@{\hskip 0.075in}c@{\hskip 0.075in}c@{}}\toprule
Model & B1 & B4 & M & RL & Cr & S & Sr \\
\midrule
Baseline\footnotemark[8] \cite{Drossos_2020_icassp} & 0.389 & 0.015 & 0.084 & 0.262 & 0.074 & 0.033 & 0.054 \\
Wav2CLIP & 0.393 & 0.054 & 0.104 & 0.271 & 0.100 & 0.045 & 0.073 \\
\bottomrule
\end{tabular}
\caption{Results of audio captioning, compared with baseline. Due to space limit, we exclude Bleu2/3, list Bleu1/4 (B1/4), METEOR (M), ROUGEL (RL), CIDEr (Cr), SPICE (S), and SPIDEr (Sr).}
\label{tab:audio_captioning}
\vspace{-1.0em}
\end{table}


On the right of Table \ref{tab:all_tasks} we have retrieval tasks. For audio retrieval (AR) task we see that Wav2CLIP as frame-level feature extractor also performs competitively. For CMR, we reach $\sim$0.05 MRR, meaning that we are able to retrieve relevant clips from top 20 on average (from 15k clips). OpenL3 performs worse, even it is also trained with audiovisual, but not forced to project to the shared embedding space. For audio captioning (AC), we show results in Table \ref{tab:audio_captioning}, we outperform baseline slightly on all metrics. We understand that it is not a fair comparison as both encoder and decoder architectures are different, but we would like to show that it is easy to adapt Wav2CLIP to different tasks, and still has reasonable performance.

\subsection{YamNet vs Wav2CLIP: Different Pretext Tasks}

\begin{figure}[ht]
\centering
\includegraphics[width=\linewidth]{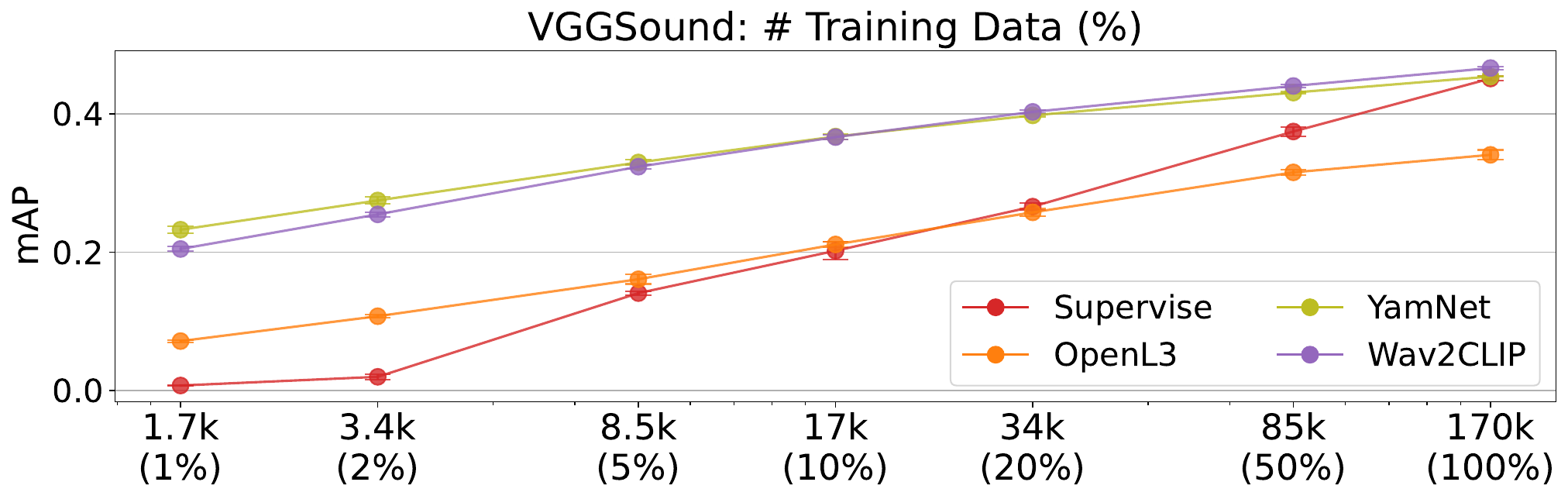}
\caption{VGGSound audio classification results with different percentage (\%) of training samples. Comparisons between Supervise, OpenL3, YamNet, and Wav2CLIP.}
\label{fig:train_num}
\vspace{-1.0em}
\end{figure}

From Figure \ref{fig:train_num}, we see results of amount of training data to train downstream classifiers on VGGSound, compared among Supervise, OpenL3, YamNet, and Wav2CLIP. Both YamNet and Wav2CLIP only require $\sim$1/10 amount of data to reach similar performance as Supervise counterpart in the smaller labeled data region, and eventually all others outperform OpenL3 when enough labels are available.

YamNet and Wav2CLIP yield very similar trend throughout all our comparisons (Table \ref{tab:all_tasks} and Figure \ref{fig:train_num}), except for FSD50K and TAU, where we hypothesize that the match between pretext and downstream tasks does affect the performance. However, this is still surprising that they perform very similarly on all other classification tasks, as the pretext tasks for pre-training are very different, while YamNet requires large amount of labeled data, Wav2CLIP is pre-trained without human annotations, and is therefore easier to scale.



\begin{figure}[htb]
\centering
\includegraphics[width=\linewidth]{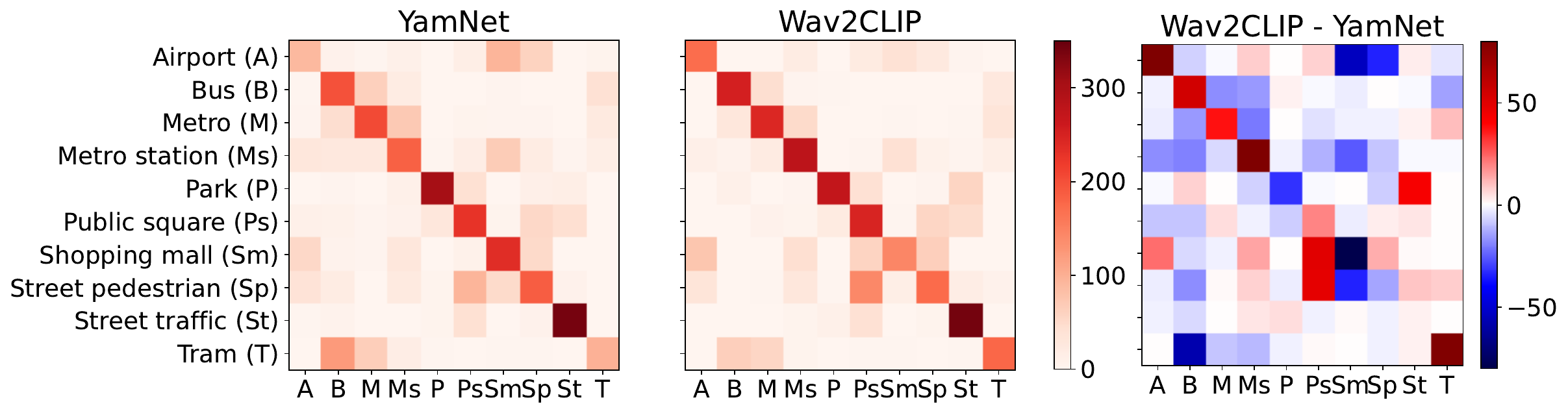}
\caption{Confusion matrices of YamNet, Wav2CLIP, and the difference between them (more red on the diagonal and blue off-diagonal the better).}
\label{fig:tau_cm}
\vspace{-1.0em}
\end{figure}


To better understand the trade-offs between pre-training on large labeled audio data, or via audio-visual correspondence, we plot the confusion matrices of YamNet and Wav2CLIP in Figure \ref{fig:tau_cm} on TAU, an audio-visual scene classification dataset. We see that Wav2CLIP can disambiguate between Airport, Shopping mall, and Street pedestrian, Tram, and Bus - these environments all have similar soundscapes but have different visual characteristics, which are distilled from the CLIP model into Wav2CLIP. However, Park and Street traffic might co-exist in the video we used for pre-training, and Wav2CLIP struggles there. On the other hand, we noticed that Wav2CLIP struggles to disambiguate between singing and speech, as the visual characteristic for both is the same (a human face). In future work, hybrid methods can be explored that leverage both sources of information for better accuracy.

\section{Image Generation: Listening at a Glance}
\label{sec:crossmodal_image_generation}

\begin{figure}[ht]
\centering
\includegraphics[width=\linewidth]{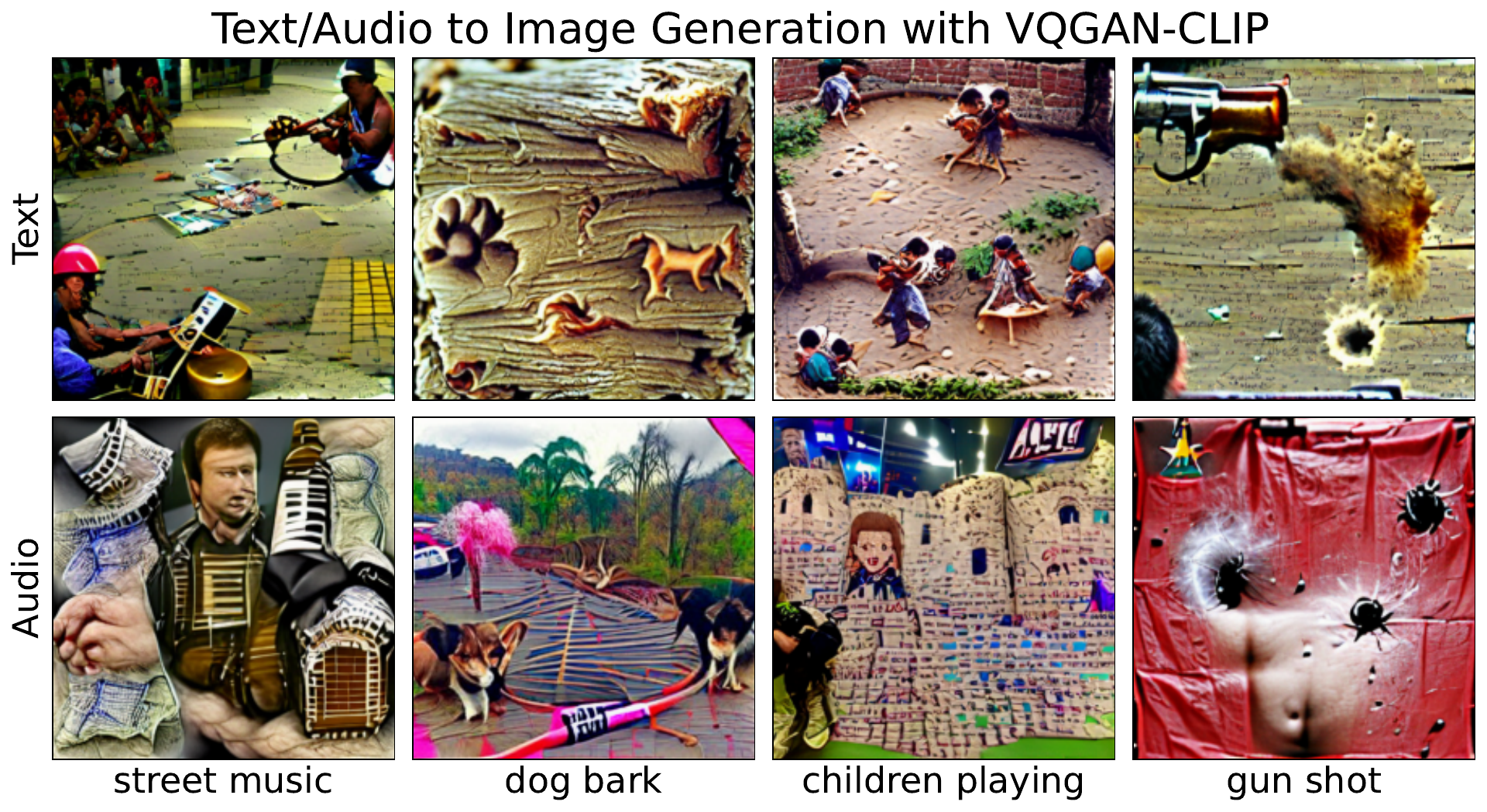}
\caption{Examples of text (top) and audio (bottom) to image generation from UrbanSound8K dataset, corresponding labels in x-axis.}
\label{fig:vqgan}
\vspace{-1.0em}
\end{figure}

\begin{figure}[ht]
\centering
\includegraphics[width=\linewidth]{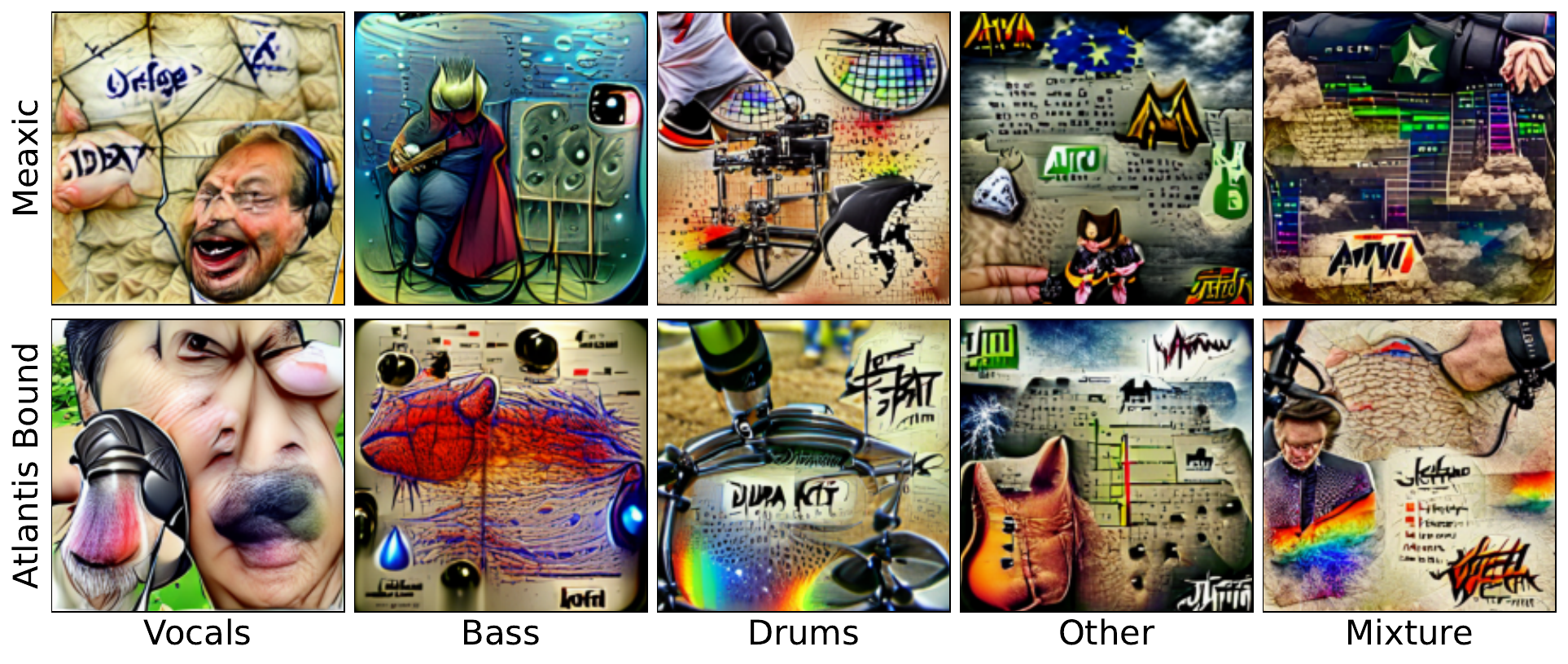}
\caption{Images generated from individual tracks and mixtures of two songs in musdb18 \cite{musdb18-hq} dataset. Top: Meaxic - Take A Step, Bottom: Atlantis Bound - It Was My Fault For Waiting.}
\label{fig:music_composition}
\vspace{-1.0em}
\end{figure}

For qualitative analysis of Wav2CLIP and the shared embedding space, we explore cross-modal image generation from audio, where we use VQGAN-CLIP, an image generation framework based on VQGAN \cite{esser2021taming}, and trained on ImageNet. CLIP embeddings are used to guide searches through latent space of VQGAN to find images that match a text prompt. We replace the input text/image embedding prompts with Wav2CLIP audio embeddings for image generation. Here are several example images generated with both text and audio prompts from UrbanSound8K in Figure \ref{fig:vqgan}. The plausible images with reasonable semantic meanings further indicates that Wav2CLIP audio embeddings are projected to a meaningful shared space. Furthermore, we show images generated from two songs of musdb18 \cite{musdb18-hq}, from each individual track, and the mixture in Figure \ref{fig:music_composition}. For the same type of instruments, diverse images are generated due to the noise introduced in the latent space with fine-tuning on each prompt. We can also observe compositionality of the embeddings, where images of the mixture contain some mixes from individual components. For more examples, please visit our project website\footnote{https://descriptinc.github.io/lyrebird-wav2clip}.


\section{Conclusion}
\label{sec:conclusion}

We propose Wav2CLIP, a method for distilling audio representations from CLIP, and evaluate systematically learned embeddings with diverse set of downstream tasks. We show that Wav2CLIP outputs general and robust audio representations and performs well across all audio classification and retrieval tasks, comparing with YamNet and OpenL3. We also demonstrate that Wav2CLIP can be easily transferred to solve multimodal tasks. For future work, we would like to experiment with various loss functions and projection layers designed specifically for multimodal data, and explore explainable machine learning methods utilizing cross-modal (audio to image) generation capabilities. Furthermore, we would like to work on audio generation from the shared embedding space, to enable the capability of cross-modal generation from text/image to audio.



\bibliographystyle{IEEEbib}
\bibliography{refs}

\end{document}